# Pseudo-Spin Based Dynamical Model for Polarisation Switching in Ferroelectrics

Zidong Wang (王子东) and Malcolm J. Grimson
*Department of Physics, the University of Auckland, Auckland 1010, New Zealand*
*E-mail address: Zidong.Wang@auckland.ac.nz*

A microscopic view of the response of the electric dipoles to a dynamic external field in a ferroelectric (FE) chain has been studied by two spin dynamics methods. One is the prominent micromagnetic approach, and the other is the micromagnetic approach with a variable size of the pseudo-spin. The energy stored in the ferroelectric chain is described by the transverse Ising model (TIM) with electric pseudo-spins. The simulations are based on a modified Landau-Lifshitz-Gilbert (LLG) equation which is precession free. The results obtained are shown and compared with the result supplemented by Landau-Devonshire (L-D) theory in the Appendix.

## I. INTRODUCTION

The study of the properties on the ferroelectric (FE) materials has been received intense theoretical and experimental investigations. On the theoretical aspect, the Landau-Devonshire (L-D) theory is used to describe the response of the electric polarisation to an external driving field in the FE material [1,2]. The L-D theory is a phenomenological theory in the mean-field approximation and it focusses on the free energy difference between the ordered and the disordered phases [3,4]. The dynamics of L-D theory has been described by a partial differential equation, named the Landau-Khalatnikov equation [2,5].

In this paper, we investigate the driven behaviour of a FE material using the microscopic viewpoint described by a spin dynamics approach. In particular, a pseudo-spin model has been used to represent the electric dipoles in the FE material. As the Fig.1 shows, an electric dipole is a pair of opposite charges separated by some distance. The electric pseudo-spin, which is beside the dipole in the figure, is a hypothetical arrow representing the dipole, it points to the positive charge, and the length of it represents the magnitude of the dipole moment. It was first introduced by P. G. de Gennes in 1963 [6], to study the phase transition in the order-disorder and the KDP-type FE crystals. The term 'pseudo-spin' was coined by R. J. Elliott, *et.al.*, in 1970 in a transverse Ising model (TIM) used to describe the energy stored in the pseudo-spin system [7,8]. Later, W. Zhong, et.al, developed a first-principles approach to study the structural phase transitions and finite temperature properties in perovskite FE materials [9,10]. This approach can be applied to BaTiO3, PbTiO3, KNbO3, etc. The novelty here is that, the system is driven by an external electric field and the time evolution of the response of the pseudo-spins is solved by a mean-field Landau-Lifshitz-Gilbert (LLG) equation [11,12]. The LLG equation is named for Lev D. Landau and Evgeny M. Lifshitz, for their contribution on the study of the precessional motion of magnetisation in 1935 [13], and T. L. Gilbert modified the equation in 1955 [14].

The results obtained by a modified LLG equation are shown in Section II, for unit size of the pseudo-spins. In Section III, the size of pseudo-spins is allowed to vary with the strength of their effective fields. A conclusion is drawn in Section IV. In the Appendix, we compare the result with mean-field L-D theory.

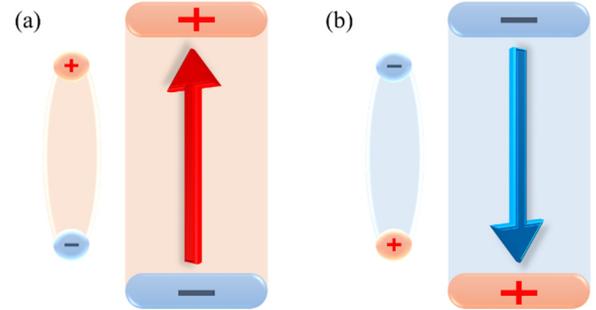

Fig. 1. Schematic of the electric dipoles and the relative electric pseudo-spins.

## II. SPIN DYNAMICS

In spin dynamics approach, we consider a 1-D FE chain consisting of $N$ sites of electric pseudo-spins. The total energy in the FE chain is described by an Ising model in a constant transverse field [15-17]. Thus, the Hamiltonian of TIM $\mathcal{H}$ is shown as,

$$\mathcal{H} = -\Omega^x \sum_{i=1}^{N} S_i^x - J \sum_{\langle i,j \rangle}^{N} \left( S_i^z S_j^z \right) - \epsilon E(t) \sum_{i=1}^{N} S_i^z \qquad (1)$$

where $S_i^x$ and $S_i^z$ are the *x*- and *z*-components of the vector $\mathbf{S}_i$ representing the pseudo-spin at site $i$. $S_i^z$ is the polarisation in the *z*-direction. In Eq. (1), the first term stands for the transverse energy with the *x*-directional transverse field $\Omega^x$, is applied perpendicular to the Ising *z*-direction [18]. The second term characterises the nearest-neighbour exchange interaction energy, with an interaction coupling $J > 0$ gives the ferroelectric phase, and the notation $\langle i, j \rangle$ represents that the sum is restricted to nearest-neighbour pairs of spins, each pair being counted only once [19]. The last term is, if the system is subject to a dynamic electric driving field $E(t)$ in the $z$-direction that couples to the pseudo-spins in the system, and $\epsilon$ is the multiplier.

The purpose of spin dynamics is to predict the time evolution of the response of the pseudo-spins to an external driving field $E(t)$. This is can be done by solving the LLG equation at the atomic level. Since in the FE material, the electric dipole moment is a measurement of the separation of positive and negative charges, it is a scalar (i.e. Fig. 1). In the spin dynamics approach, the time evolutions of the pseudo-spins are expected to perform a precession free trajectory. Therefore, the dynamic responses of electric pseudo-spins are expressed as [17,20-23],

$$\frac{\partial \mathbf{S}_i}{\partial t} = -\lambda \left[ \mathbf{S}_i \times \left( \mathbf{S}_i \times \mathbf{H}_i^{eff} \right) \right] \quad (2)$$

where $\lambda$ is the intrinsic damping parameter, and $\mathbf{S}_i = (S_i^x, S_i^y, S_i^z)$ represent the consecutive pseudo-spin with three directional components for a unit size of $\|\mathbf{S}_i\| = 1$. $\mathbf{H}_i^{eff}$ is the effective field for each pseudo-spin, which is defined as a functional derivative of Eq. (1) [24],

$$\mathbf{H}_i^{eff} = -\frac{\delta \mathcal{H}}{\delta \mathbf{S}_i} = \begin{bmatrix} \Omega^x \\ 0 \\ JS_j^z + \epsilon E(t) \end{bmatrix} \quad (3)$$

It is notable that the electric effective field in Eq. (3), without any $y$-component, and only has a constant transverse field in the $x$-component.

To demonstrate the response of the pseudo-spins to an external electric field using the spin dynamics approach. We found that the maximal size tractable in reasonable CPU time is given by $N = 100$. However, in this theoretical study, we introduce the normalised variables, i.e. $\mathbf{s}_i = \mathbf{S}_i / S_0$, $t' = \lambda t$ and $\mathbf{h}_{S_i}^{eff} = -(1/\epsilon)\delta \mathcal{H}/\delta \mathbf{s}_i$. The other parameters with a '*' label are characterised dimensionless. Thus, the damping parameter is normalised to $\lambda^* = 1$. The coupling of the nearest-neighbour exchange interaction is set as $J^* = 8$. The transverse field is $\Omega^* = 10$. The external field is set in a sinusoidal form, i.e. $E(t) = E_0 \sin(\omega t)$, with fixed amplitude of $E_0^* = 10$ in the $z$-direction only, and a switching frequency $\omega^* = 0.1$. The multiplier is normalised to $\epsilon^* = 1$. Periodic boundary conditions have been applied, and the initial states have been set at random.

The numerical results show that the mean magnitudes of the pseudo-spins in Fig. 2, obtained by a fourth-order Runge-Kutta method with a time step $\Delta t = 0.0001$. Mean responses of pseudo-spins in the $x$-, $y$- and $z$-directions against the time are shown in Fig. 2(a). The figure shows the response of the $x$-component $<S_x>$ (*green dotted curve*) is driven by the transverse field, and against the magnitude of the $z$-component. The magnitude of the $y$-component $<S_y>$ (*blue dashed curve*) is small, due to the absence of a driving field applied along the $y$-direction. Hence, the response in the $z$-direction $<S_z>$ (*black solid curve*), i.e. the polarisation, dominates the motion of the pseudo-spin, and a familiar hysteresis loop is plotted in Fig. 2(b). This hysteresis loop agrees well with theory, except for the magnitude of the saturation values are always confined within the unit size of pseudo-spin, more discussions are drawn in Sections III. To investigate the spin behaviour in detail, a plot of the trajectory of one bulk pseudo-spin from the FE chain is presented in Fig. 2(c) [17,22]. The trajectory of the pseudo-spin has a shell shape and, for the sake of clarity, the multiple colours characterise the different magnitudes of the $z$-component $S_z$. The $S_z$ reduces the magnitude following by the strength of driving field, it can allow a growth of $S_x$ to catch up the fixed size of the pseudo-spin (i.e. $\|\mathbf{S}_i\| = 1 = \sqrt{S_x^2 + S_y^2 + S_z^2}$). Inversely, the $S_x$ can decrease due to the $S_z$ gains the magnitude. These features also can be observed in Fig. 2(a).

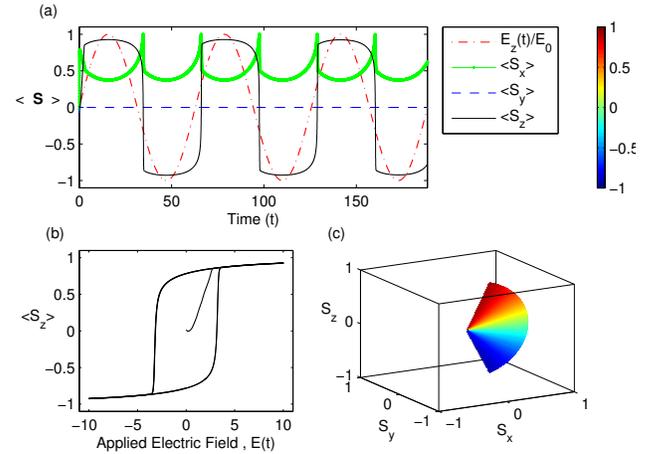

Fig. 2. Simulated results by spin dynamics approach. (a) Time evolution of mean responses in the $x$-, $y$-, and $z$-directions of the pseudo-spins to an applied external field. (b) Hysteresis loop of the $z$-component mean response. (c) Spin trajectory of a pseudo-spin, multiple colours characterise the magnitudes in the $z$-direction. '$<>$' on the $y$-labels and legends represent a mean value.

## III. SPIN DYNAMICS WITH VARIABLE-SIZE PSEUDO-SPIN

Actually, the polarisation is defined as the electric dipole moment density. For a homogeneous dielectric material, the polarisation $P$ is proportional to an applied electric field $E_{ext}$, as [4]

$$P = \epsilon_0 \chi_e E_{ext} \quad (4)$$

where $\epsilon_0$ is the electric permittivity of free space. The susceptibility $\chi_e$ can be determined from Eq. (4) by the derivative of polarisation to applied field, such as [24,25]

$$\chi_e = \frac{1}{\epsilon_0} \frac{\delta P}{\delta E_{ext}} \quad (5)$$

Similar to the spin dynamics case, the polarisation is represented by magnitude of the pseudo-spin in each direction together with their effective fields. Therefore Eq. (5) can be re-written as

$$\aleph = \frac{1}{\epsilon_0} \left\| \frac{\delta S_i}{\delta H_i^{eff}(t)} \right\| \quad (6)$$

where $\aleph$ is the dimensionless susceptibility in the spin dynamics system. Corresponding to Eq. (6), the size of the pseudo-spin is proportional to its effective field. Since the effective field is obtained by Eq. (3), it can be varied with the dynamic external field. Therefore, the pseudo-spin should have a variable size. To relax the limitation of the unit size of the pseudo-spin, we introduce the variable size of pseudo-spin as '$\boldsymbol{\xi}_i$', to distinguish with the unit size of pseudo-spin '$S_i$'. According to Eq. (6), the length of variable size pseudo-spin $\boldsymbol{\xi}_i = (\xi_i^x, \xi_i^y, \xi_i^z)$ can be deduced as,

$$\|\boldsymbol{\xi}_i(t)\| = \epsilon_0 \aleph \|H_i^{eff}(t)\| \quad (7)$$

The new TIM Hamiltonian and the precession free LLG equation for the variable-size pseudo-spins are given in Eq. (8) and Eq. (9), respectively.

$$\mathcal{H} = -\Omega^x \sum_{i=1}^N \xi_i^x - J \sum_{\langle i,j \rangle}^N \left( \xi_i^z \xi_j^z \right) - \epsilon E(t) \sum_{i=1}^N \xi_i^z \quad (8)$$

$$\frac{\partial \boldsymbol{\xi}_i}{\partial t} = -\lambda \left[ \boldsymbol{\xi}_i \times \left( \boldsymbol{\xi}_i \times H_i^{eff} \right) \right] \quad (9)$$

We use the same dimensionless parameters and conditions as Section II. The susceptibility is set as $\aleph = \left( \Omega^{*2} + E_0^{*2} \right)^{-1/2}$. The results are shown in Fig. 3 with same details as Fig. 2. In Fig. 3(a), the maximum magnitude of the z-component $\xi_i^z$ is higher. The gradient of the wings to the hysteresis loop in Fig. 3(b) is larger than unit size pseudo-spin in Fig. 2(b). This leads to a more complex structure to the trajectory of the pseudo-spin in Fig. 3(c). We infer that is a result of the relaxation delays the response time of the z-component $\xi_i^z$ to the driving field, hence the x-component $\xi_i^x$ trends to fill reinforcement the rest size of the pseudo-spin. This behaviour also depicted in Fig. 3(a). Generally, the response of the x-component $\xi_i^x$ (*green dotted line*) keeps constantly. Unless, as the $\xi_i^z$ starts to change the phase, by shrinking its size and passing through the origin. Consequently, the $\xi_i^x$ can jump up a bit due to it is supported by a constantly transverse field.

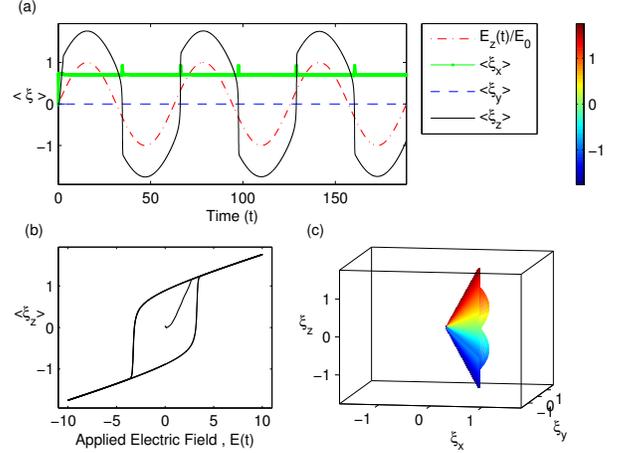

Fig. 3. Simulated results of the variable-size pseudo-spins by spin dynamics approach. The details are as for Fig. 2.

## IV. CONCLUSION

The main point of this paper is to observe the response of polarisation in a FE material to a driving field in a mean-field microscopic model. The pseudo-spin model has been used to represent the electric dipoles in the spin dynamics approach with a TIM Hamiltonian to describe the total free energy and its time dependence. Importantly, as the external field increases, the pseudo-spin size increases and the polarisation grows. The numerical results utilise a precession free LLG equation and demonstrate that the response of the pseudo-spins is consistent with the theory. A comparison with the result of the phenomenological L-D theory in the Appendix, shows excellent agreement. However, while the L-D theory is a phenomenological, the spin dynamics approach is appropriate for a microscopic study.

## APPENDIX: LANDAU-DEVONSHIRE THEORY

To show the feasibility of the results in the spin dynamics approach. We demonstrate the familiar Landau theory calculation. The FE chain is considered as alignment of the electric dipoles. The energy of the chain can be expressed by the L-D free energy $\mathcal{F}(P;T;E)$ for second-order phase transition, in terms of the external field $E(t)$ and polarisation $P_i$ are shown in Eq. (10) [5].

$$\mathcal{F} = \sum_{i=1}^N \left[ \frac{1}{2}\alpha P_i^2 + \frac{1}{4}\beta P_i^4 - \epsilon E(t) P_i + \frac{1}{2}\kappa (P_{i+1} - P_i)^2 \right] \quad (10)$$

The phenomenological investigations of dynamical free energy can be governed by the Landau-Khalatnikov equation with

$$\frac{\partial P_i}{\partial t} = -\frac{1}{\tau}\frac{\delta F}{\delta P_i}$$
$$= -\frac{1}{\tau}\left[\alpha P_i + \beta P_i^3 - \epsilon E(t) + \kappa(P_{i+1} + P_{i-1} - 2P_i)\right] \quad (11)$$

where $\tau$ is the relaxation time. The result from this equation is shown in Fig. 4, which is identical to Fig. 3(b).

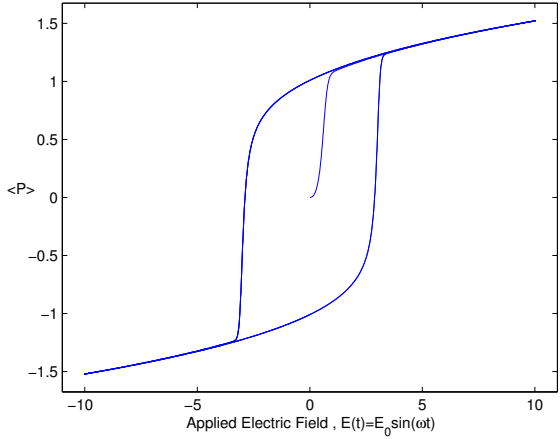

Fig. 4. The hysteresis loop simulated by first-derivative L-K equation with $\alpha^* = -5$, $\beta^* = 5$ and $\kappa^* = 8$. '<>' on the y-label represents a mean polarisation.

## ACKNOWLEDGEMENTS

Z. D. Wang gratefully acknowledges Wang YuHua (王玉华) and Zhao BingJin (赵秉金) for financial support.